\documentclass[11pt]{article}

\usepackage{geometry}

\usepackage{lineno}

\usepackage{setspace}

\usepackage{graphicx}

\usepackage{amsmath}

\begin{document}

\title{Antibiotic treatment, duration of infectiousness, and disease transmission}

\maketitle

\begin{center}

\author{Thomas Caraco$^{}\footnote{E-mail: tbcaraco@gmail.com}$}

\end{center}

\begin{center}
\emph{Department of Biological Sciences, University at Albany, Albany NY 12222, USA}
\end{center}


\begin{spacing}{1.8}

\begin{flushleft}

\parindent=4mm
\emph{Abstract}.
Humans, domestic animals, orchard crops, and ornamental plants are commonly treated with antibiotics in response to bacterial infection.  By curing infectious individuals, antibiotic therapy might limit the spread of contagious disease among hosts.  But an antibiotic`s suppression of within-host pathogen density might also reduce the probability that the host is otherwise removed from infectious status prior to therapeutic recovery.  When rates of both recovery \emph{via} treatment and other removal events (\emph{e.g}., isolation or mortality) depend directly on within-host pathogen density, antibiotic treatment can relax the overall removal rate sufficiently to increase between-host disease transmission.  To explore this dependence, a deterministic within-host dynamics drives the infectious host's time-dependent probability of disease transmission, as well as the probabilistic duration of the infectious period.  At the within-host scale, the model varies (1) inoculum size, (2) bacterial self-regulation, (3) the time between infection and initiation of therapy, and (4) antibiotic efficacy.  At the between-host scale the model varies (5) the size/susceptibility of groups randomly encountered by an infectious host.  Results identify conditions where antibiotic treatment can increase duration of a host`s infectiousness, and consequently increase the expected number of new infections.  At lower antibiotic efficacy, treatment might convert a rare, serious bacterial disease into a common, but treatable infection.

\vspace{4mm}


\section{Introduction}
\label{intro}
Antibiotics are administered routinely to humans, agricultural/pet animals, and certain plants \cite{McManus_2002,Dagata_2008,Gualerzi_2013}.  The most common objective of antibiotic treatment is therapeutic control of an individual`s bacterial infection \cite{Levin_2014}.  Beyond concerns about the evolution of resistance \cite{Read_2011,Lopatkin_2017}, use of antibiotics to treat infection presents challenging questions, including optimizing trade-offs between antibacterial efficacy and toxicity to the treated host \cite{Geli_2012}.  This study asks if antibiotic treatment of an infection can have untoward consequences at the population scale; the paper models an antibiotic's direct impact on within-host pathogen dynamics and resulting, indirect affects on between-host transmission \cite{Mideo_2008,Childs_2019}.

The model assumes that antibiotic therapy can reduce within-host pathogen density sufficiently to cure the host`s infection.  But the antibiotic`s suppression of bacterial density extends the average waiting time for the host`s removal from infectiousness via other processes (\emph{e.g}., isolation, hospitalization or mortality).  The paper`s focal question asks how varying the age of infection when antibiotic treatment begins impacts both the duration of disease and the intensity of transmission during the host`s infectious period.  When removal equates with mortality from disease, the results identify conditions under which an antibiotic may simultaneously increase both survival of an infected individual and the expected number of secondary infections.

\subsection{The infectious period}
Efficacious antibiotics, by definition, reduce within-host pathogen density \cite{Levin_2010}; for some infections, antibiotics consequently increase host survival.  Therapeutic recovery of a treated individual may imply an epidemiological benefit.  If antibiotics shorten the infectious period, the count of infections per infection could decline \cite{Levin_2014}.  This interpretation follows from SIR compartment models, where neither the host-removal rate nor the antibiotically-induced recovery rate depends explicitly on within-host pathogen density.  That is, antibiotics are assumed to reduce duration of the infectious period and to exert no effect on per-individual transmission intensity.  By extension, antibiotics may then reduce pathogen transmission.

However, antibiotic therapy might, in other cases, increase the expected length of the infectious period.  Relationships among transitions in host status must often depend on a within-host dynamics \cite{Gilchrist_2002,Mideo_2008}.  As infection progresses, the pathogen density's trajectory should drive change in the rate of host removal while ill (\emph{e.g}., isolation), the rate of recovery from disease, as well as the rate at which infection is transmitted \cite{Reluga_2010,Vanderwaal_2016}.  For many human bacterial infections, an individual can still transmit the pathogen after beginning antibiotic therapy \cite{Moon_2019}.  Common infections remain transmissible for a few days to two weeks \cite{Siegel_2007}; although not addressed here, sexually transmitted disease may persist within a host for months after antibiotic therapy has begun \cite{Falk_2015}.  Therapeutic reduction in pathogen density might eventually cure the host, while allowing the host to avoid hospitalization, isolation, \emph{etc} during treatment \cite{derigne_2016}.  The result can be a longer period of infectious contacts and, consequently, increased secondary infections.

This paper assumes that with or without antibiotic treatment, a diseased host`s infectious period may be ended by a removal process that depends on within-host pathogen density.   As a convenience, removal includes any event terminating infectious contacts with susceptible hosts, prior to the antibiotic curing the disease.  Social isolation \cite{Huffman_1997}, hospitalization for humans, and host mortality will be more or less probable removal events for any particular disease.  But they are equivalent in that they end the infectious period.  The model assumes that an antibiotic, by deterring within-host pathogen growth, increases the expected waiting time for removal, but an increase in antibiotic efficacy reduces the time elapsing until the host is cured.  This interaction affects the count of secondary infections; disease reproduction numbers (before and after therapy begins) identify conditions where an antibiotic increases the spread of disease.

\subsection{Random encounters: susceptible groups}

When infection is rare, random variation in the number of contacts between diseased and susceptible hosts influences whether or not the pathogen spreads at the population scale \cite{Bailey_1964,vanbaalen_2002}.  Therefore, this paper treats reproduction numbers, \emph{i.e}., infections per infection, as random variables \cite{Antia_2003}.  Social group size can govern contacts between infectious and susceptible hosts, and so affect transmission of new infections \cite{Brown_2001,Turner_2008,Caraco_2016}.  The model below asks how the number of hosts per encounter with an infectious individual (with the product of encounter rate and group size fixed) impacts the variance in the count of secondary infections; specifically, the paper asks how group size impacts the probability that a rare infection fails to invade a host population \cite{Caraco_2014,Lahodny_2015}.

\subsection{Organization}
The model treats within-host pathogen dynamics and its antibiotic regulation deterministically \cite{Dagata_2008}.  Removal from the infectious state and between-host transmission are modeled probabilistically \cite{Whittle_1955,Caillaud_2013,Lindberg_2018}.

At the within-host scale, the model considers both density-independent and self-regulated pathogen growth.  The host`s removal rate and the infection-transmission intensity will depend directly on the time-dependent bacterial density.  Pathogen density increases monotonically from time of infection until antibiotic treatment begins, given persistence of the host`s infectious state.  The antibiotic then reduces pathogen density until the host is cured or removed prior to completing therapy (whichever occurs first).

Counts of secondary infections will require the temporal distribution of infectious contacts, since the probability of transmission depends on the time-dependent pathogen density \cite{Strachan_2005,Vanderwaal_2016}.  The results explore effects of antibiotics and inoculum size \cite{Steinmeyer_2010} on length of the infectious period, disease reproduction numbers, and pathogen extinction.

\section{Within-host dynamics: timing of antibiotic treatment}
\label{within}
For many bacterial infections of vertebrates, little is known about within-host pathogen growth \cite{Haugen_2019}.  In a laboratory system, \emph{Pseudomonas aeruginosa} infection of \emph{Drosophila melanogaster} \cite{Mulcahy_2011}, the pathogen increases exponentially until either the host dies or antibacterial treatment begins \cite{Dargenio_2001,Heo_2009,Lindberg_2018}.  In more complex host-pathogen systems, resource limitation or physical crowding must often decelerate pathogen growth within the host, implying self-regulation \cite{Ebert_1997,Austin_1998,Oloughlin_2013,Ankomah_2014}.  Numerical results plotted below compare ways in which the strength of self-regulation interacts with an antibiotic to influence duration of infectiousness, and intensity of pathogen transmission.

$B_t$ represents the within-host bacterial density at time $t$; $B_0$ is the inoculum size.  Antibiotic treatment begins at time $t_A > 0$.  Table \ref{symbols} defines model symbols used in this paper.

If the pathogen grows exponentially prior to treatment, $B_t = B_0 e^{r t}$ for $t \leq t_A$.  The intrinsic growth rate $r > 0$ is the difference between bacterial replication and mortality rates per unit density.  The latter rate may reflect a nonspecific host immune response \cite{Pilyugin_2000}; the model does not include an explicit immune dynamics, to focus on effects of antibiotic timing and efficacy.  Under logistic self-regulation, the per-unit growth rate becomes $(r - c B_t)$, where $c$ represents intraspecific competition.  For this case, the within-host density prior to treatment becomes:
\begin{equation}
B_t = r{\bigg{/}}\left[c + \left(\frac{r}{B_0} - c\right) e^{-r t} \right];~~t \leq t_A \nonumber
\end{equation}
where $B_0 < r/c$; the inoculum should be smaller than the ``carrying capacity.''  For the same $(B_0, r)$, the self-regulated density cannot, of course, ever exceed the exponentially growing density between time of infection and initiation of antibiotic therapy.  For both growth assumptions, $B_{t_A}$ represents the within-host density at initiation of antibiotic therapy.

\begin{table}[t]
\centering
\begin{tabular}{|c|l|}
\hline
Symbols & Definitions \\
\hline
  \textbf{Within-host scale} & ~ \\
  $t$ & Time since infection (hence, age of infection) \\
  $B_t$ & Bacterial density at time $t$ after infection, pathogen state  \\
  $B_0$ & Inoculum size \\
  $r$ & Pathogen`s intrinsic rate of increase  \\
  $c$ & Pathogen intraspecific competition   \\
  $\gamma_A^*$ & Density-independent bacterial mortality rate due to antibiotic \\
  $t_A$ & Age of infection when antibiotic initiated \\
  $\theta$ & Proportionality of inoculum to pathogen density at time of cure \\
  $t_C$ & Age of infection when host cured\\
      \hline
  \textbf{Individual host scale} & ~ \\
  $h_t$ & Removal rate of host infective at time $t$  \\
  $\phi$ & Removal-rate prefactor\\
  $\eta$ & Infection-severity parameter \\
  $L_t$ & Probability host remains infectious at time $t \leq t_C$  \\
      \hline
  \textbf{Between-host scale} &  ~  \\
  $\lambda/G$ & Stochastic contact rate, group of $G$ susceptibles $(G = 1, 2, ... )$ \\
  $\nu_t$ & Conditional probability of infection, given contact \\
  $\xi$ & Infection susceptibility parameter  \\
  $p_t$ & Probability susceptible infected at time $t$; $p_t = L_t \nu_t$ \\
  $\mathcal{P}_j$ & Time-averaged probability of infection at contact \\
  ~ & ~~~~~~~~~~before/after $(j = 1, 2)$ therapy begins \\
  $R_1$ & Expected new infections per infection before $t_A$ \\
  $R_2$ & Expected new infections per infection on $(t_A, ~t_C)$ \\
  $R_0$ & $R_1 + R_2$ \\
  $\mathcal{B}_{0j}$ & Inoculum transmitted, before/after $(j = 1, 2)$ therapy begins \\
  \hline
\end{tabular}
\caption{Definitions of model symbols, organized by scale.}
\label{symbols}
\end{table}

Most antibiotics increase bacterial mortality \cite{Regoes_2004,Levin_2010}, though some impede replication \cite{Austin_1998}.  When a growing bacterial population is treated with an efficacious antibiotic, bacterial density (at least initially) declines exponentially \cite{Tuomanen_1986,Balaban_2004,Wiuff_2005}.  Hence, the model below assumes that a bactericidal antibiotic causes exponential decay of $B_t$ during therapy.  The Discussion acknowledges complications that might arise during treatment.

\subsection{Host states}
The host becomes infectious at time $t = 0$, and remains infectious until either removed or cured by the antibiotic.  That is, no secondary infections occur after removal or therapeutic cure, whichever occurs first.  Hence, transmission can occur during antibiotic therapy, prior to cure.  If the host remains infectious at time $t$, both the probability of disease transmission (given encounter with a susceptible) and the removal rate depend explicitly on within-host density $B_t$.

\subsection{Antibiotic concentration and efficacy}
Assumptions concerning antibiotic efficacy follow from Austin et al. [1998].  Given that the host remains infectious at time $t > t_A$, the total loss rate per unit bacterial density is $\mu + \gamma (A_t)$, where $A_t$ is plasma concentration of antibiotic, and $\gamma$ maps $A_t$ to bacterial mortality per unit density.

Assume that the antibiotic is effectively `dripped' at rate $D_A$.  Plasma antibiotic concentration decays through both metabolism and excretion; let $k_A$ represent the total decay rate. Then, $dA_t /dt = D_A - k A_t$, so that $A_t = \left(D_A/k\right)~(1 - e^{- kt})$, for $~t > t_A$.  Antibiotic concentration generally approaches equilibrium faster than the dynamics of bacterial growth or decline \cite{Austin_1998}.  Then a quasi-steady state assumption implies the equilibrium plasma concentration of the antibiotic is $A^{*} = D_A/k$.

Bacterial mortality increases in a decelerating manner as antibiotic concentration increases \cite{Mueller_2004,Regoes_2004}.  Using a standard formulation \cite{Geli_2012}:
\begin{equation}
\gamma (A_t) = \Gamma_{max} ~A_t/\left(a_{1/2} + A_t\right);~~~t > t_A
\end{equation}
where $\gamma (A_t) = \Gamma_{max}/2$ when $A_t = a_{1/2}$.  Applying the quasi-steady state assumption, let $\gamma_A^* = \gamma(A^*)$.  Since the antibiotic is efficacious, $\gamma_A^* > r$.  If antibiotic concentration cannot be treated as a fast variable, time-dependent analysis of concentration is available \cite{Austin_1998}.

\subsection{Antibiotic treatment duration}
Antibiotic therapy begins at time $t_A$.  During treatment, within-host pathogen density declines as $dB_t/dt = - \left( \gamma_A^* - r\right) B_t$.  Then:
\begin{equation}
\label{Bafter}
B_t ~= ~B_{t_A} ~exp \left[ - (\gamma_A^* - r) (t - t_A)\right];~~~t > t_A;~\gamma_A^* > r
\end{equation}
\noindent
where only $B_{t_A}$ depends on the presence/absence of self-regulation.  For the exponential case $B_t = B_0 ~exp \left[ rt - \gamma_A^* (t - t_A) \right]$ after treatment begins.  For self-regulated pathogen growth:
\begin{equation}
B_t = \left[ \frac{r e^{\gamma_A^* t_A}}{c e^{r t_A} + \left( \frac{r}{B_0} - c \right)} \right] e^{- (\gamma_A^* - r) t}
\end{equation}
Specific bacterial densities, but not the form of antibiotically produced decline, depend on the absence/strength of self-regulation.

Given that the host is not otherwise removed, antibiotic treatment continues until the host is cured at time $t_C > t_A$.  A `cure' means that the within-host pathogen density has declined sufficiently that the host no longer can transmit the pathogen; a cure need not imply complete clearance of infection.  $t_C$ is the maximal age of infection; that is, no host remains infectiousness beyond $t_C$.  In terms of pathogen density, $B(t_C) = B_0/\theta$, where $\theta \geq 1$.  For exponential pathogen growth, we have:
\begin{equation}
\label{tc}
B_0/\theta = B_0~exp\left[ rt_C - \gamma_A^* (t_C - t_A)\right]~~~~\Rightarrow ~~~~t_C = \frac{\gamma_A^* t_A + ln \theta}{\gamma_A^*  - r} > t_A
\end{equation}
\noindent
If the cure requires only that $B_t$ return to the inoculum size, then $\theta = 1$, and $t_C  = \gamma_A^* t_A /( \gamma_A^*  - r ) > t_A$.  Instead of defining recovery \emph{via} therapy as a pathogen density proportional to $B_0$, suppose that the host is cured if the within-host density declines to $B(t > t_A) = \tilde{B} \leq B_0$.  Let $\tilde{\theta} = \tilde{B}/B_0$.  The associated maximal age of infection is $\tilde{t} = (\gamma_A^* t_A - ln \tilde{\theta})/(\gamma_A^* - r)$.  $\tilde{t}$ depends on $\gamma_A^*$, $t_A$ and $r$ just as $t_C$ does, and numerical differences will be small unless $B_0$ and $\tilde{B}$ differ greatly.

If pathogen growth self-regulates and $B(t_C) = B_0/\theta$, the host is cured at:
\begin{equation}
\label{tclog}
t_C = \frac{\gamma_A^* t_A + ln \theta}{\gamma_A^* - r} - ln \left[ 1 + \frac{B_0}{r/c} \left( e^{r t_A} - 1\right) \right] (\gamma_A^* - r)^{-1}
\end{equation}
\noindent
Since $B(t_A)$ is smaller under self-regulated growth than under density independent growth, the antibiotic cures the host faster under self-regulation.  The seemingly counterintuitive effect of $B_0$ in Eq. \ref{tclog} occurs because $B(t_C)$ is proportional to $B_0$.  For both exponential and self-regulated pathogen growth, any $\theta \geq 1$ implies that $t_C$ declines as $\gamma_A^*$ increases.

\section{Duration of infectious state}
\label{host}
Removal includes any event, other than antibiotic cure, that ends the host's infectious period.  Removal occurs probabilistically; importantly, the instantaneous rate of removal depends on pathogen density.  Noting that removal by mortality becomes more likely with the severity of ``pathogen burden'' \cite{Medzhitov_2012}, the model assumes that the removal rate at any time $t$ strictly increases with pathogen density $B_t$.

The model takes removal as the first event of a nonhomogeneous Poisson process; $h_t$ is the instantaneous rate of removal at time $t$ \cite{Bury_1975}.  $L_t$ is the probability that the host, infected at time 0, remains infectious at time $t \leq t_C$.  Prior to initiation of therapy:
\begin{equation}
L_t \equiv exp\left[ -\int_0^t h_{\tau} ~d\tau\right];~~~t \leq t_A
\end{equation}
and $(1 - L_t)$ is the probability the host has been removed before time $t$.  $h_t$ is the stochastic removal rate at time $t$; let $h_t = \phi B_t^{\eta};~~~\phi,~\eta > 0$.  The parameter $\phi$ scales bacterial density to the timescale of removal.  Removal is more/less likely as bacterial density increases/decreases.  For $t \leq t_A$, $h_t$ has the form of the Gompertz model for age-dependent mortality among adult humans \cite{Missov_2013}.  $h_t$ assumes that the likelihood of removal saturates $(\eta < 1)$, increases linearly $(\eta = 1)$, or accelerates $(\eta > 1)$ with increasing pathogen density, depending on the particular host-pathogen combination.

Suppose the pathogen grows exponentially before time $t_A$.  Then the host remains infectious prior to antibiotic treatment with probability:
\begin{equation}
\label{Lexp}
L_t = exp \left[ - \phi B_0^{\eta} \int_0^t e^{\eta r \tau} d \tau \right] = exp \left[ \frac{\phi B_0^{\eta}}{\eta r}\right]{\bigg{/}} ~exp\left[ \frac{\phi B_t^{\eta}}{\eta r}\right];~~~t \leq t_A
\end{equation}
where the numerator is a positive constant, and the denominator strictly increases before the antibiotic begins.  Equivalently, $L(t < t_A)$ under exponential growth can be written:
\begin{equation}
\label{L1}
L_t =  exp\left[ - \frac{\phi}{\eta r} \left( B_t^{\eta} - B_0^{\eta}\right)\right];~~~t \leq t_A
\end{equation}
$L(t = 0) = 1$, and persistence of infection declines as $t$ increases.

For logistic pathogen growth prior to treatment, we have:
\begin{equation}
\label{L1sr}
L_t = exp \left[ - \phi r^{\eta} \int_{0}^{t} ~\frac{d \tau}{\left[ c + \left( \frac{r}{B_0} - c \right) e^{-r \tau}\right]^{\eta}}\right]
\end{equation}
\noindent
For given $(B_0, r)$, the self-regulated density at $t \in (0,~t_A)$ must be lower than the unregulated density.  Consequently, $L_t$ under exponential growth cannot exceed the corresponding probability when pathogen growth self-regulates.  If removal equates with host mortality, self-regulated pathogen dynamics increase the chance that the host survives until antibiotic treatment begins.

\subsection{Antibiotic therapy: removal \emph{vs} cure}
\label{Lantibiotic}
If an infectious host begins antibiotic therapy, the individual must have avoided removal through time $t_A$.  During antibiotic treatment, a host has instantaneous removal rate:
\begin{equation}
\label{h2}
h_t = \phi B_{t_A}^{\eta} ~e^{- \eta (\gamma_A^* - r) (t - t_A)};~~~t > t_A
\end{equation}
\noindent
where, again, only $B_{t_A}$ depends on the presence/absence of self-regulation.  The probability that the host remains infectious at any time $t$, where $t_A < t < t_C$, is the probability of entering treatment in the infectious state, $L_{t_A}$, times the probability of avoiding removal from $t_A$ to $t$, given the host`s state at $t_A$.  Using Eq. (\ref{h2}), the probability that the host remains infectious during treatment is:
\begin{equation}
L_t = L_{t_A} ~exp\left[- \phi B_{t_A}^{\eta}~\int_{t_A}^t e^{- \eta (\gamma_A^* -r)(\tau - t_A)} d\tau \right];~~~t > t_A
\end{equation}
where $B_{t_A}$ and, consequently, $L_{t_A}$ depend on presence/absence of self-regulation.

Using $B(t > t_A)$ as given by Eq. \ref{Bafter}, we have the probability that infectiousness persists to time $t$ during therapy, for either presence or absence of self-regulation prior to therapy:
\begin{equation}
L_t = L_{t_A} ~ exp \left[ \frac{\phi B_t^{\eta}}{\eta (\gamma_A^* - r)}\right]{\bigg{/}}exp\left[ \frac{\phi B_{t_A}^{\eta}}{\eta (\gamma_A^* - r)}\right]= L_{t_A} ~exp \left[- \frac{\phi}{\eta (\gamma_A^* - r)} \left( B_{t_A}^{\eta} - B_t^{\eta} \right)\right];~t > t_A
\end{equation}
where $B_{t_A} > B_t$, and $L_{t_A}$ is given by either Eq. \ref{L1} or Eq. \ref{L1sr}, as appropriate.

\begin{figure}[t]
  \centering
\vspace*{6.0truecm}
    \includegraphics{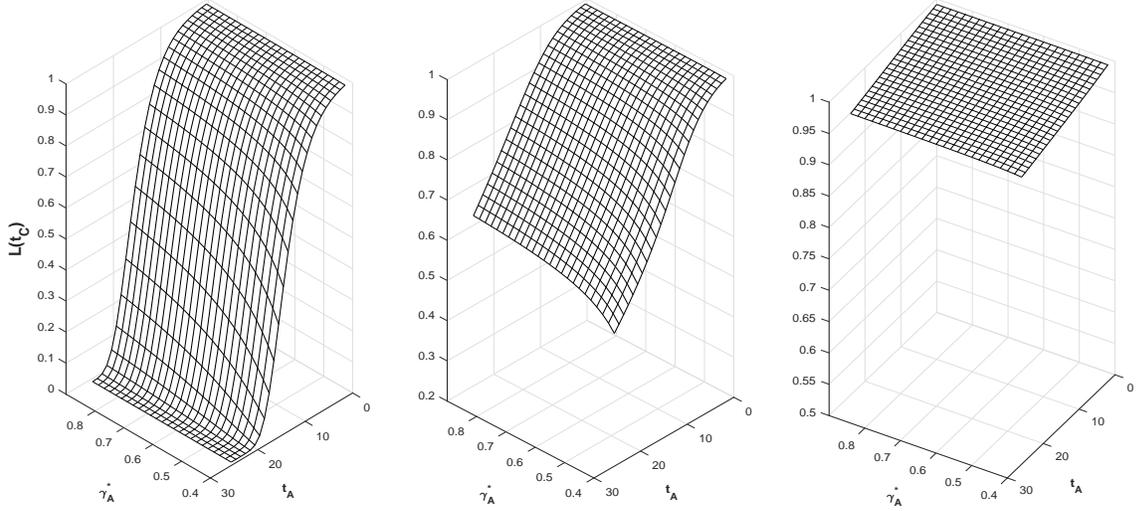}
\vspace*{2.0truecm}
\caption{Probability antibiotic cures host.  Equivalently, probability host remains infectious (avoids removal) until $t_C$.  \textbf{Left plot}:  Exponential within-host dynamics.   $L(t_C)$ declines rapidly as $t_A$ increases.  For most levels of $t_A$, greater antibiotic efficacy $\gamma_A^*$ increases probability host will be cured.  \textbf{Middle plot}: Self-regulation; $c = 10^{-6}$.  $L_(t_C)$ intermediate between density independence and very strong self-regulation. \textbf{Right plot}: Strong self-regulation; $c = 10^{-4.8}$.  Greater self-regulation increases chance antibiotic cures host (note $z$ axis scale), difference maximal at greater $t_A$ and low antibiotic efficacy.  All plots: $B_0 = 10^4$, $r = 0.3$, $\phi = 10^{-7}$, $\eta = \theta = 1.0$.  Note that since $B(t_C)$ is fixed, while $B_{t_A}$ varies inversely with strength of self-regulation, $t_C$ declines as self-regulation increases.
}
\label{survive}
\end{figure}
The infectiousness-survival probabilities $L_t$ collect some direct and indirect consequences of model assumptions.  Delaying initiation of therapy (\emph{i.e}., increasing $t_A$) increases $B_{t_A}$ and hence must decrease $L_{t_A}$, the probability that infectiousness persists until treatment begins.  Rephrased, delaying antibiotic therapy increases the chance that the host is removed (and so stops transmitting infection) before therapy begins.  Since $\partial B_{t_A}/\partial t_A > 0$, the time required for the antibiotic to cure the host $(t_C - t_A)$ must increase with $t_A$.  Since increased $t_A$ decreases $L_{t_A}$ and increases  $(t_C - t_A)$, then $\partial L(t_C)/\partial t_A < 0$; the probability that the remains infectious until cured decreases with delayed initiation of treatment.  These effects always hold for exponential growth, and hold for logistic growth as long as $B_{t_A} < r/c$, the carrying capacity.

Increasing bacterial self-regulation moderates, but does not reverse, these effects of $t_A$.  For $r/c$ large enough, $B_{t_A}$ declines as $c$ increases.  Then $L_{t_A}$ must increase, and $(t_C - t_A)$ must decrease.  Hence $\partial L(t_C)/\partial c \geq 0$; the probability that the host is cured therapeutically never decreases with stronger self-regulation.

Greater antibiotic efficacy (increased $\gamma_A^*$) does not affect $B_{t_A}$ or $L_{t_A}$.  Intuitively, $t_C$ declines, and $L(t_C)$ is non-decreasing, as $\gamma_A^*$ increases.  Surfaces in Fig. \ref{survive} show $L(t_C)$, the probability that the host remains infectious until cured at $t_C$; strength of bacterial self-regulation increases from the left plot to the right.  In each plot $L(t_C)$ declines as the delay prior to antibiotic treatment increases.  Greater antibiotic efficacy increases $L(t_C)$ across most levels of $t_A$, when growth is exponential.  The effect diminishes as bacterial self-regulation increases.  Clearly, the likelihood the host is cured increases for most $(t_A,~\gamma_A^*)$ combinations as the strength of self-regulation increases.  Greater self-regulation reduces $B_{t_A}$, but does not affect $\theta B_0$, the density where the host is cured.  Since antibiotic efficacy also is independent of the level of self-regulation, $t_C$ is reduced as self-regulation grows stronger.

The model's simple within-host pathogen dynamics allows the rate of removal and, by a complementarity, persistence of the infectious state to depend on within-host pathogen density.  Proceeding, the time-dependent probability of infection transmission will also depend on within-host density \cite{Ganusov_2003}.

\section{Transmission}
\label{transmit}
The focal infective contacts susceptible hosts as groups.  Each group has the same size $G$; often $G = 1$.  Contacts occur as a Poisson process, with constant probabilistic rate $\lambda /G$; the contact rate does not depend on time or pathogen state $B_t$.  Then the expected number of individuals contacted in any period does not  depend on susceptible-host group size $G$.

Given that the host remains infectious and a transmission-contact occurs at time $t$, associate a random, dichotomous outcome $I_t(j)$ with susceptible host $j$; $j = 1, 2, ..., G$.  $I_t(j) = 0$ if no transmission occurs, and $I_t(j) = 1$ if a new infection occurs, independently of all other contact outcomes.  A contact, then, equates to $G$ independent Bernoulli trials, and the number of new infections, \emph{per} contact, follows a binomial probability function with parameters $G$ and $p_t$.  $p_t = \textrm{Pr}[I_t = 1]$, the conditional probability that any host $j$ acquires the infection, given contact at time $t$.  The model writes $p_t$ as a product: $p_t = L_t \nu_t$, the unconditional probability that the host remains infectious through time $t$.  $L_t$, weighs ``births'' of new infections upon contact \cite{Ganusov_2003,Day_2011}.  $\nu_t$ is the conditional probability that any host $j$ is infected at time $t$ given that the host remains infectious at time $t$, and contact occurs.  Both $L_t$ and $\nu_t$ depend on within-host pathogen density $B_t$.

Given an encounter, the transmission probability $\nu_t$ assumes a dose-response relationship \cite{Strachan_2005,Kaitala_2017}.  Following a preferred model \cite{Tenuis_1996}, $\nu_t = 1 - exp [- \xi B_t]$, where $\xi$ is the susceptibility parameter.  Then $p_t = L_t (1 - e^{- \xi B_t})$.  $\nu_t$ decelerates with $B_t$ since infection of a single host saturates with propagule number \cite{Keeling_1999,vanbaalen_2002,Caraco_2006}.  Note that $\partial \nu_t/\partial B_t > 0$, and $\partial h_t/\partial B_T > 0$.  An increase in the transmission probability, due to greater within-host pathogen density, is constrained by a greater removal rate.

\subsection{New-infection probabilities: before and during treatment}
\label{R0}
New infections occur randomly, independently both before and after treatment begins.  Since $\textrm{d}B_t/\textrm{d}t$ changes sign at $t_A$, let $R_1$ represent the expected number of new infections on $(0,~t_A]$; let $R_2$ be the expected number of new infections on $(t_A,~t_C]$.  For simplicity, refer to these respective time intervals as the first and second period.  $R_0$ is the expected total number of new infections \emph{per} infection; $R_0 = R_1 + R_2$.

From above, encounters with the infectious host occur as a Poisson, hence memoryless, process.  Suppose that $N$ such encounters occur on some time interval $(t_x,~t_y)$.  By the memoryless property, the times of the encounters (as unordered random variables) are distributed uniformly and independently over $(t_x,~t_y)$ \cite{Ross_1983}.  Uniformity identifies the time averaging for the conditional infection probability $p_t$.  For the first period, the unconditional (\emph{i.e}., averaged across the initial $t_A$ time periods) probability of infection at contact is $\mathcal{P}_1$:
\begin{equation}
\label{bigP1}
\mathcal{P}_1 = \frac{1}{t_A}~\int_0^{t_A} p_{\tau} ~\textrm{d} \tau = \frac{1}{t_A}~ \int_0^{t_A} L_{\tau} (1 - e^{- \xi B_{\tau}}) ~\textrm{d} \tau
\end{equation}

Eq. \ref{bigP1} applies to both exponential and self-regulated growth.  For the former case, we have:
\begin{equation}
\mathcal{P}_1 = \left( exp \left[ \frac{\phi B_0}{\eta r}\right]{\bigg{/}}t_A\right) ~\left( \int_0^{t_A} exp\left[ - \frac{\phi B_{\tau}^{\eta}}{\eta r}\right] \textrm{d} \tau -  \int_0^{t_A} exp\left[ - \frac{\phi B_{\tau}^{\eta}}{\eta r}  - \xi B_{\tau}\right] \textrm{d} \tau \right)
\end{equation}
where $B_{\tau} = B_0 e^{r \tau}$.

For the second period, averaging uniformly yields $\mathcal{P}_2$, the averaged infection probability after treatment begins.  $\mathcal{P}_2$ has the same form as Eq. \ref{bigP1}, with averaging over the time period $(t_C - t_A)$, and bounds on integration changed accordingly.  For both exponential and logistic within-host growth, we obtain:
\begin{equation}
\label{bigP2}
\mathcal{P}_2 = \left( \frac{L_{t_A}}{t_C - t_A}{\bigg{/}}exp \left[ \frac{\phi B_{t_A}^{\eta}}{\eta (\gamma_A^* - r)}\right]\right)~\left( \int_{t_A}^{t_C} exp \left[ \frac{\phi B_{\tau}^{\eta}}{\eta (\gamma_A^* - r)}\right] \textrm{d} \tau - \int_{t_A}^{t_C} exp \left[ \frac{\phi B_{\tau}^{\eta}}{\eta (\gamma_A^* - r)} - \xi B_{\tau} \right] \textrm{d} \tau \right)
\end{equation}
where $L_{t_A}$ is given above (and depends on the presence/absence of self-regulation), and $B_{\tau}$ is given by Eq. (\ref{Bafter}).  Biologically, $\mathcal{P}_1$ and $\mathcal{P}_2$ collect effects of within-host density, modulated by antibiotic treatment, on between-host transmission of infection.  Since the within-host dynamics affects both persistence of the infectious state and the probability of transmitting infection upon contact, the strength of self-regulation should impact the number of secondary infections \emph{per} infection.

\subsection{$R_0$}
For each of the two periods, the number of infections sums a random number of random variables.  Each element of the sum is a binomial variable with expectation $G\mathcal{P}_z$ and variance $G\mathcal{P}_z (1 - \mathcal{P}_z)$, where $z = 1,~2$.  The number of encounters with susceptible hosts is a Poisson random variable with expectation and variance during the first period $(\lambda /G) t_A$, and expectation during the second period $(\lambda /G) (t_C - t_A)$.  Let $X_1$ be the random count of new infections during the first period, and let $X_2$ be the second-period count.  Then from the time of infection until antibiotic treatment begins, $R_1 = E[X_1] = \lambda \mathcal{P}_1 t_A$, and $V[X_1] = R_1 [1 + P_1 (G - 1)]$.  Then, $R_2 = E[X_2] = \lambda \mathcal{P}_2 (t_C - t_A)$, and the variance of $X_2$ is $R_2 [1 + P_2 (G - 1)]$.  Note that if $G = 1$, each $X_z$ is Poisson with equality of expectation and variance.  By construction, the expected number of infections both before and after antibiotic treatment begins does not depend on group size $G$.  But each variance of the number of new infections increases with group size.  Finally, the total number of new infections \emph{per} infection has expectation $R_0$, where $R_0 = E[X_1 + X_2] = \lambda \left[ \mathcal{P}_1 t_A + \mathcal{P}_2 (t_C - t_A)\right]$.  The variance of the total number of new infections is $V[X_1 + X_2] = R_0 + (G - 1)[\mathcal{P}_1 R_1 + \mathcal{P}_2 R_2]$.

Since group size affects only the variance of the reproduction numbers, any increase in group size can increase $\textrm{Pr}[X_1 + X_2 = 0]$, the probability of no new infections, even though $R_0 > 1$.  No new infections requires that each $X_z = 0$; $z = 1, ~2$.  The probability of no pathogen transmission at a single encounter is $(1 - \mathcal{P}_z)^G$, since outcomes for the $G$ susceptible hosts are mutually independent.  Given $n$ encounters in period $z$, the conditional probability of no new infections during that period is $\textrm{Pr}[X_z = 0 \mid n] = [(1 - \mathcal{P}_z)^G]^n$.  Then, unconditionally:
\begin{equation}
\label{PrExtinct}
\textrm{Pr}[X_z = 0] = \sum_{n=0}^\infty [(1 - \mathcal{P}_z)^G]^n \textrm{Pr}[n]
\end{equation}
Since $(1 - \mathcal{P}_z)^G < 1$, $\textrm{Pr}[X_z = 0]$ is given by the probability generating function for $n$, evaluated at $(1 - \mathcal{P}_z)^G$.  From above, $n$ is Poisson with parameter $(\lambda/G) t_A$ during the first period.  Then:
\begin{equation}
\label{pgf}
\textrm{Pr}[X_1 = 0] = exp \left[ (\lambda/G) t_A ([1 - \mathcal{P}_1]^G - 1)\right]
\end{equation}
For the second period, $\textrm{Pr}[X_2 = 0] = exp \left[ (\lambda/G) (t_C - t_A) ([1 - \mathcal{P}_2]^G - 1)\right]$.  Each $\textrm{Pr}[X_z = 0]$ increases as $G$ increases; the group-size effect is stronger as the infection probability $\mathcal{P}_z$ increases.  The probability that no new infection occurs is, of course, the product of the independent probabilities.

\section{Numerical Results}
\label{plots}
Plots in Figs. \ref{RzLt} and \ref{extinctG} show results motivating this paper.  Consider first how $R_0$ varies with $t_A$, at different levels of self-regulation.  If antibiotic therapy begins immediately after the host becomes infectious ($t_A < 2$ in Fig 2c) then $R_0 < 1$ in both the presence and absence of self-regulation; the disease will fail to invade a susceptible population.  But antibiotics are seldom administered at the onset of infectiousness \cite{Gualerzi_2013}.  Delaying therapy a bit ($2 < t_A < 4$) allows the infection to spread.  That is, reasonably rapid initiation of antibiotic treatment allows $R_0 > 1$, for both self-regulated and unregulated growth before $t_A$.  As $t_A$ continues to increase, only strong self-regulation produces further, though quickly decelerating, increase in $R_0$.  More interestingly, for both exponential growth and weak self-regulation, delaying therapy sufficiently ($t_A > 9$) leaves $R_0 < 1$ again, inhibiting the spread of infection among hosts.  For the exponential example, results for larger $t_A$ equate essentially to no antibiotic therapy: ($L_{t_A} \rightarrow 0; R_2 = 0$).  In these cases relatively early initiation of antibiotic therapy increases the probability the host will be cured (Fig. 2f), but allows the disease to advance among hosts ($R_0 > 1$).  But no antibiotic therapy (or $t_A$ delayed sufficiently) prevents initial spread of infection ($R_0 < 1$).

Why does increasing the time elapsing between infection and initial treatment (or no treatment) sometimes reduce the chance that disease will spread?  Why does relaxed pathogen self-regulation increase this effect?  A small $t_A$ implies a low $B_{t_A}$; early treatment maintains a reduced within-host density and a consequently reduced removal rate for $t > t_A$.  The host's chance of being cured, rather than first being removed, increases when treatment begins relatively soon after infection.  That is, therapy begun at low $t_A$ more likely cures the host, but (on average) leaves the host infectious longer.  The latter effect maximizes $R_0$ at a lower $t_A$ in both the exponential and weakly self-regulated examples.

Earlier initiation of treatment must reduce $R_1$.  For exponential and weakly self-regulated pathogen growth, the spread of infection among hosts, for low $t_A$, is due more to transmission during antibiotic treatment; $R_0$ and $R_2$ each reaches its maximum at nearly the same $t_A$ value.  For any $t$, $t_A < t < t_C$, the reduction in the infection probability $\nu_t$ due to the antibiotic's regulation of within-host pathogen density is more than compensated by the increase in $L_t$, the probability that the host remains infectious.  The focal point is that $R_0 < 1$ with no antibiotic therapy, though $R_0$ can exceed $1$ with therapy.  When removal and therapeutic cure without removal depend differently on the within-host dynamics, this non-obvious effect of $t_A$ can occur.

Stronger self-regulation reduces $B_{t_A}$ and consequently lowers the removal rate for $t > t_A$.  The host is then more likely to remain infectious until cured .  The example with strong self-regulation reduces the time-dependent removal rate enough that $R_0$ increases monotonically with increasing $t_A$.

\begin{figure}[t]
  \centering
\vspace*{6.3truecm}
    \includegraphics{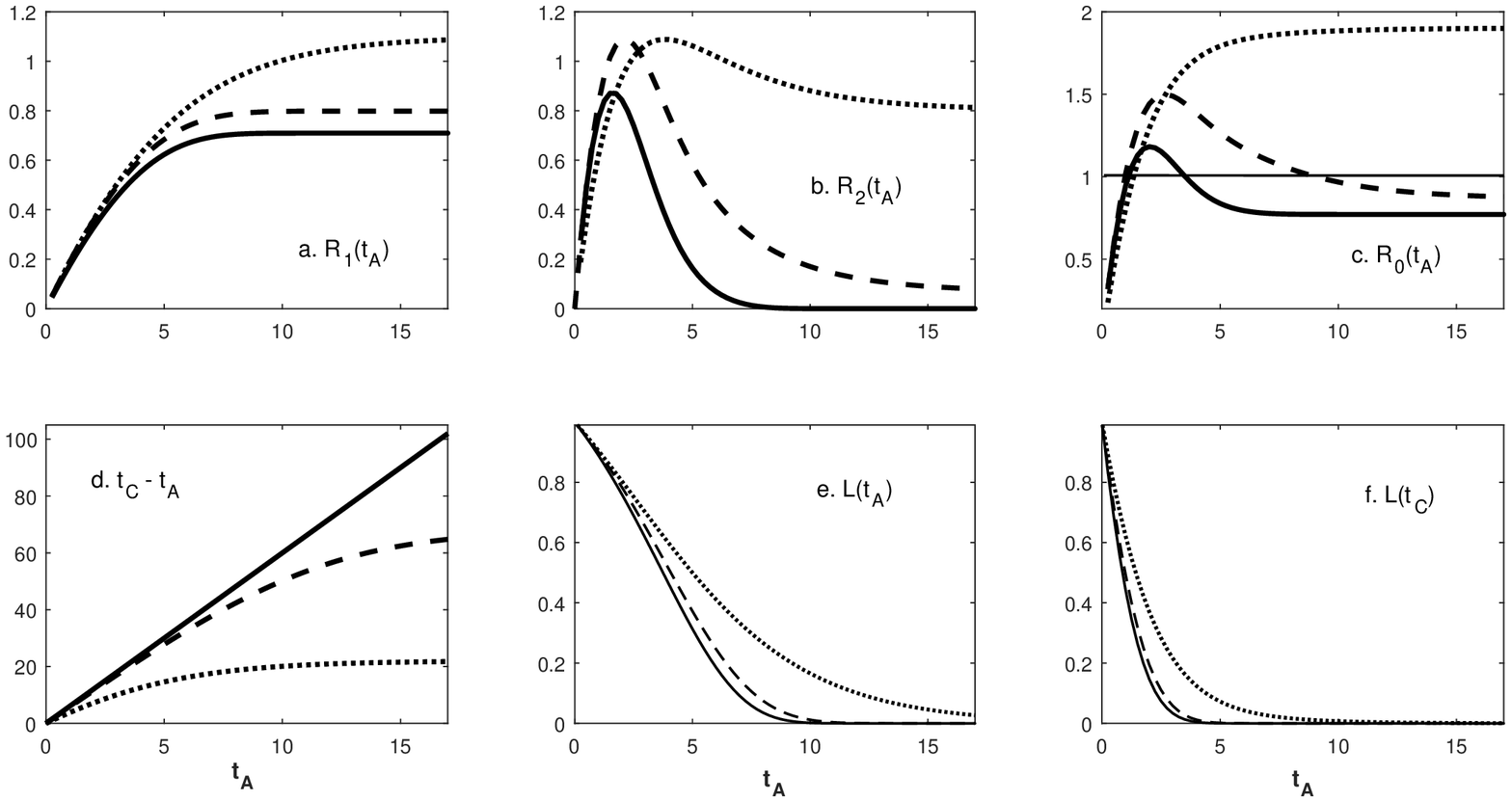}
\vspace*{2.0truecm}
\caption{Early antibiotic therapy can promote infection transmission.  Each plot: solid line is exponential, dashed line is weak self-regulation ($c = 10^{-6}$), dotted line is stronger self-regulation ($c = 10^{-5}$).  \textbf{a}. Left: $R_1$ expected infections before $t_A$. \textbf{b}. $R_2$ expected infections after antibiotic started.   \textbf{c}. $R_0$, expected infections \emph{per} infection. \textbf{d}. Time required for treatment curing the host.  \textbf{e}.  Probability host begins antibiotic therapy.  \textbf{f}. Probability that the host is cured before removal, given therapy initiated at $t_A$.   All plots: $B_0 = 10^4$, $r = 0.3$, $\phi = 10^{-5}$, $\gamma_A^* = 0.35$, $\eta = \theta = 1.0$, $\lambda = 0.2$, $\xi = 1.0$.
}
\label{RzLt}
\end{figure}
Consider the exponential case, and suppose that avoiding removal through the antibiotic treatment implies surviving disease; the host is either removed by mortality or cured by the antibiotic.  Then, the infected host obviously benefits from therapy.  But there can be a significant cost at the among-host scale as the infection spreads.  A rare ($R_0 < 1$), but virulent infection in the absence of antibiotics can become a common ($R_0 > 1$), through treatable disease when antibiotic therapy begins soon after initial infection.

\begin{figure}[t]
  \centering
\vspace*{6.3truecm}
    \includegraphics{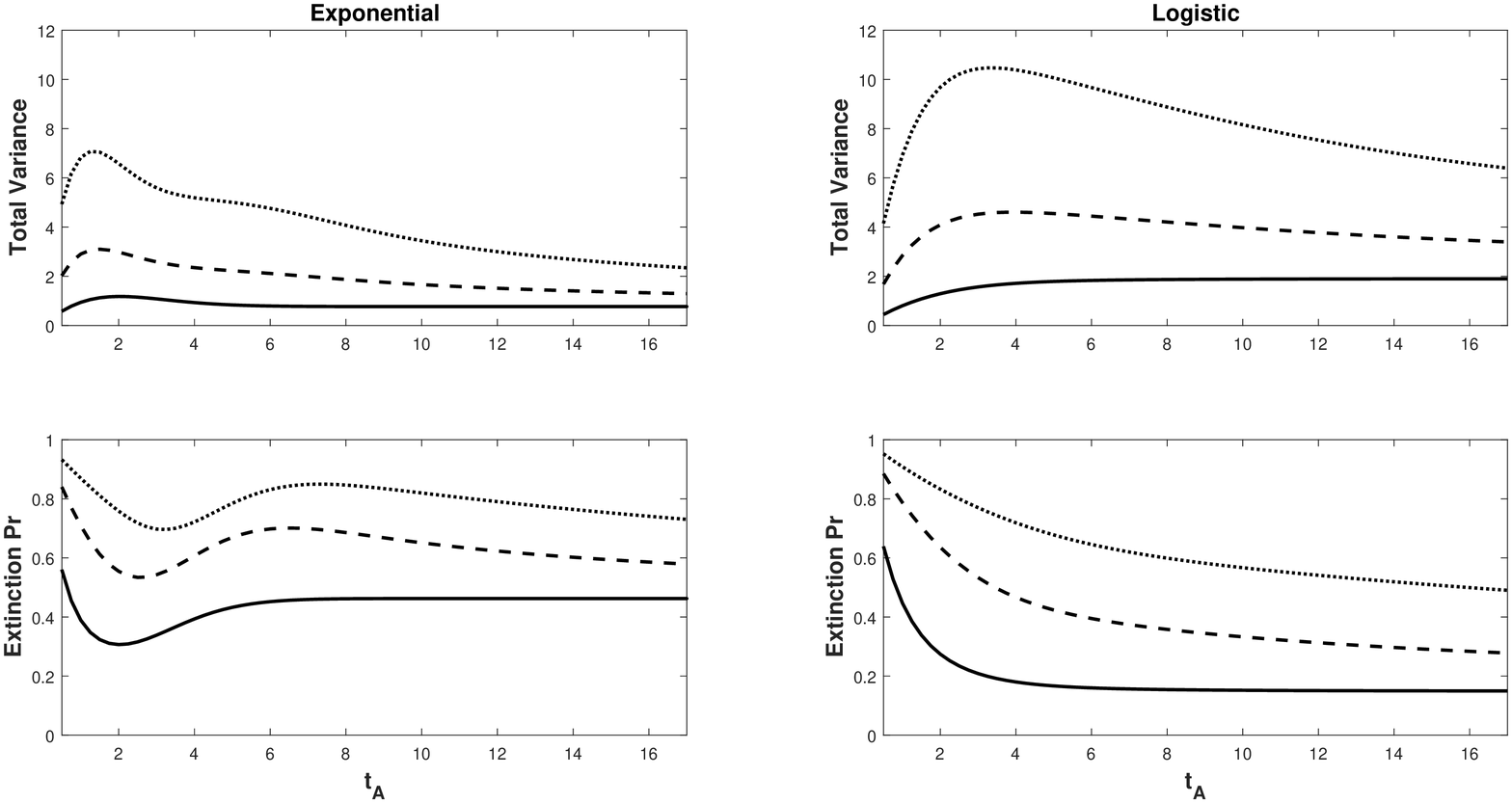}
\vspace*{2.0truecm}
\caption{Susceptible group size and disease-extinction probability.  Each plot: $G = 1$ (solid line), $G = 4$ (dashed line), $G = 10$ (dotted line).  Left column: Exponential pathogen growth, right column: logistic growth ($c = 10^{-5}$).  Variance in the random count of infections \emph{per} infection increases as susceptible groups are larger, but encountered less often.  Probability of no secondary infections \emph{per} infection (``extinction'') increases as group size increases.
}
\label{extinctG}
\end{figure}
Figure \ref{extinctG} verifies how increasing susceptible-host group size reduces the probability of at least one secondary infection, despite independence of $R_0$ and group size $G$.  Larger groups increase the variance in the total count of infections \emph{per} infection.  As a consequence, the probability that no new infections occur (pathogen ``extinction'') increases strongly with $G$.  Through mutual dependence on $t_A$, extinction is less probable as $R_0$ increases.  However, even for the $t_A$ levels maximizing $R_0$ in Fig. \ref{RzLt}, sufficiently large group size (under both exponential and weakly self-regulated growth) assures that pathogen extinction is more likely than is spread of infection.

\subsection{Inoculum size, antibiotic efficacy, and $R_0$}
\label{RofBg}
The preceding numerical results varied $t_A$, and held both inoculum size $B_0$ and antibiotic efficacy ($\gamma_A^* - r$) constant.  Variation in inoculum size can impact within-host pathogen growth \cite{Schmid_2007,White_2012}, any host immune response \cite{Gama_2012}, and host infectiousness \cite{Chu_2004,Steinmeyer_2010}.  That is, inoculum size, through effects on within-host processes, should influence infection transmission.  Of course, increasing ($\gamma_A^* - r$) should increase the likelihood of curing, rather than removing, the host.

Fig. \ref{RzB0g} simultaneously varies the inoculum $B_0$ and antibiotic mortality $\gamma_A^*$.  Dependent quantities are $R_0$ and the probability that a host remains infectious until cured [$L{t_C}$]; results were calculated for a smaller and larger $t_A$.  For given parameter values, $R_0$ reaches a maximum at low antibiotic efficacy and small inoculum size; the pattern holds for both exponential pathogen growth (subplot \textbf{a}) and stronger self-regulation (subplot \textbf{e}).  Subplots \textbf{a} and \textbf{e} show results for $t_A = 4$; the surfaces has the same shape for both smaller and larger $t_A$ levels.  Not surprisingly, $R_0$ always decreases as $\gamma_A^*$ increases, for both exponential and logistic growth.  Note that the effect of increased efficacy, observed for these parameters, does not mean that antibiotics always deter the spread of infection.

Why does $R_0$ decline as inoculum size increases?  Any increase in $B_0$ increases $B_t$ for all $t \leq t_C$.  The removal rate $h_t$ increases as a consequence, and the expected duration of infectiousness must consequently decline.  For these parameters, where susceptibility $\xi$ is comparatively large, any increase in the transmission probability $\nu_t$ with $B_t$ does not compensate for the reduction in duration of infectiousness.  Hence, by increasing the likelihood of early removal, a larger inoculum can decrease the expected number of secondary infections.  Increasing antibiotic efficacy decreases not only $R_0$, but also the sensitivity of $R_0$ to variation in inoculum size.

Subplots \textbf{c} and \textbf{d} of Fig. \ref{RzB0g} verify, for exponential growth, that the chance of the host remaining infectious until cured (\emph{i.e}., avoiding removal) declines as $B_0$ increases.  Note the clear quantitative differences between the two $L_(t_C)$-surfaces.  For any $(B_0, ~\gamma_A^*)$-combination, the host's probability of remaining infectious is greater for low $t_A$ (subplot \textbf{c}) than for high $t_A$ (subplot \textbf{d}).  If removal equates with mortality, so that $L(t_C)$ becomes the host`s survival probability, the host should `prefer' earlier initiation of therapy.

\begin{figure}[t]
  \centering
\vspace*{6.5truecm}
    \includegraphics{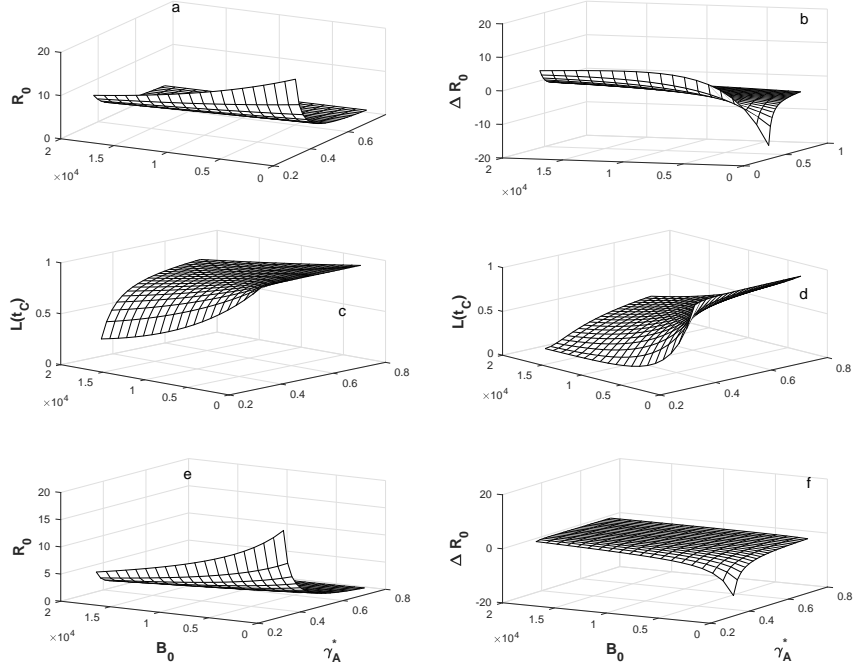}
\vspace*{2.4truecm}
\caption{Effects of varying $B_0$ and $\gamma_A^*$.  \textbf{a}. $R_0$ for $t_A = 4$; $B_t$ exponential.  $R_0$ declines monotonically as inoculum size $B_0$ increases; $R_0$ also declines as antibiotic efficacy increases.  \textbf{b}. $\Delta R_0$ is $R_0$ for $t_A = 4$ \emph{minus} $R_0$ for $t_A = 8$; exponential growth.  For medium and larger $B_0$, combined with lower antibiotic efficacy, earlier treatment generates more secondary infections.  \textbf{c}. Probability treated host is cured, $t_A = 4$.  \textbf{d}. Probability host cured, $t_A = 8$. \textbf{e}. $R_0$ for $t_A = 4$; $B_t$ logistic.  \textbf{f}. $\Delta R_0$ is $R_0$ for $t_A = 4$ \emph{minus} $R_0$ for $t_A = 8$; $B_t$ logistic.  All plots: $r = 0.3$, $\phi= 10^{-6}$, $\theta = \eta = 1$, $\xi = 0.7$, and $\lambda = 0.4$. Plots e and f: $c = 10^{-5}$, stronger self-regulation.
}
\label{RzB0g}
\end{figure}

Subplot \textbf{b} in Fig. \ref{RzB0g} shows the \emph{difference} between $R_0$ values for the two $t_A$ levels; $\Delta R_0 = R_0(t_A = 4) - R_0(t_A = 8)$.  When the antibiotic has greater efficacy ($\gamma_A^* \geq 0.5$) $\Delta R_0 \leq 0$.  A stronger antibiotic allows earlier start of therapy to decrease the expected number of secondary infections, although it extends the duration of infection (by reducing the removal rate).  However, if the antibiotic has lower efficacy ($\gamma_A^* \leq 0.4$), $\Delta R_0 > 0$ for sufficiently large $B_0$.  Earlier treatment still increases duration of the infective state (\emph{i.e}., $L(t_C)$ increases), and now also increases $R_0$.  When small inoculum size is combined with lower antibiotic efficacy, the infected host benefits most, in terms of the chance of being cured, from earlier therapy (low $t_A$).  However, the consequence is accelerated spread of infection at the among-host scale, since $\Delta R_0 > 0$.  Sufficiently strong self-regulation of within-host growth can eliminate this effect (subplot \textbf{f}), since host survival until cured is, overall, much less sensitive to variation in $t_A$ (see Fig. \ref{survive}).

\subsection{Group size, $R_0$ and pathogen `extinction'}
\label{XRG}

Fig. \ref{ExRintx} shows, for exponential pathogen growth, how varying both $R_0$ and susceptible-group size $G$ affects the probability that the focal host transmits no secondary infections.  $R_0$ was varied by computing extinction probabilities over different levels of $B_0$.  Fixing $G$ in any of the plots, pathogen-extinction probability never increases, and sometimes declines, as $R_0$ increases.  The decline in pathogen-extinction probability with increasing $R_0$ is greatest when susceptible hosts are encountered as solitaries, \emph{i.e}., when the infection-number variance is minimal.  Given $R_0$, the chance of pathogen extinction increases strictly monotonically as $G$ increases; see Eq. (\ref{pgf}).  The rate at which extinction probability increases with $G$ grows larger as $R_0$ increases.  Each plot in Fig. \ref{ExRintx} includes regions where, for sufficiently large group size, $R_0 > 1$ but pathogen extinction is more likely than not.  For logistic growth, the surfaces (not shown) are qualitatively similar.  Quantitative differences, where they occur, contract the surfaces for the logistic with respect to the $R_0$ axis.

\begin{figure}[t]
  \centering
\vspace*{8.4truecm}
    \includegraphics{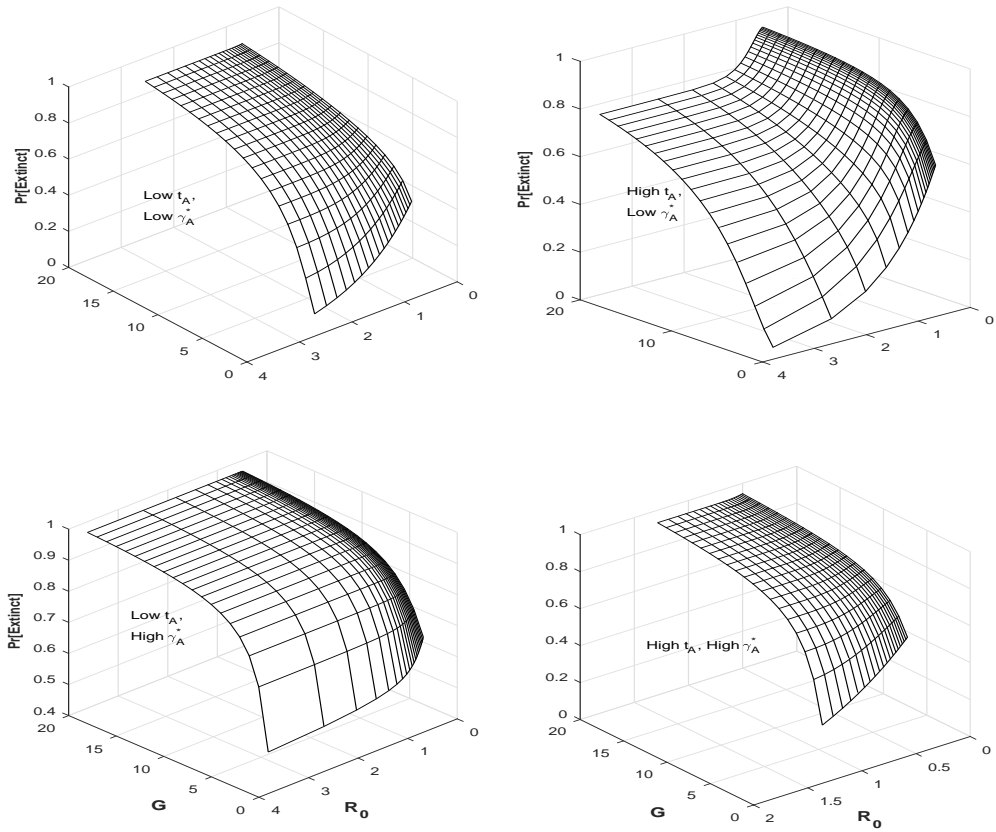}
\vspace*{2.3truecm}
\caption{Probability of no secondary infections.  Each plot shows $(Pr[X_1 = 0]~Pr[X_2 = 0])$ as a bivariate function of $R_0$ and group size $G$; note directions of axes.  $R_0$ varied by varying inoculum size $B_0$ form $10^3$ to $ 2 \times 10^4$.  \textbf{Top row}: $\gamma_A^*$ = 0.35. \textbf{Bottom row}: $\gamma_A^*$ = 0.7.  \textbf{Left column}: $t_A = 4$.  \textbf{Right column}: $t_A = 8$.  Pathogen extinction less likely as $R_0$ increases; pathogen extinction always more likely as group size $G$ increases. Increase in extinction due to larger group size increases at greater $R_0$.  Each plot shows a substantial region where $R_0 > 1$, but probability of pathogen extinction exceeds 0.5.   All plots: $r = 0.3$, $\phi= 10^{-5.5}$, $\theta = \eta = 1$, $\xi = 0.5$, $\lambda = 0.1$.
}
\label{ExRintx}
\end{figure}

The rows in Fig. \ref{ExRintx} differ in $\gamma_A^*$; the columns differ in $t_A$.  The average $R_0$ value, calculated as a function of $B_0$, ranged from 1.1 to 1.3 among the subplots; values suggesting spread of infection among hosts.  However, the average probability of no secondary infections ranged from $0.66$ to $0.75$; \emph{i.e}., exceeded $1/2$.  The model`s grouped susceptibles do not promote ``superspread,''  if the stochastic rate of such encounters is inversely proportional to group size.

\section{Discussion}
\label{discuss}
Objectives of antibiotic treatment include prophylaxis and promoting growth of agricultural animals, as well as control of individual bacterial infections; antibiotics present both scientific and societal issues \cite{Read_2011,Levin_2014}.  This paper assumes that any increase in within-host pathogen density makes removal/mortality due to infection more likely.  Antibiotic therapy reduces pathogen density and so lowers the instantaneous removal rate, until treatment cures the host and ends the infectious period.  Removal and the course of therapy interact through their separate relationships with pathogen density, and the interaction governs the expected duration of infectiousness and disease-transmission probabilities during the infectious period.

The model was motivated by two sets of observations.  First, adults and children routinely take antibiotics (often accompanied by fever-reducing medicine) for upper respiratory infections, and then return to work or school as soon as symptoms begin to subside.  In some cases these presentees \cite{Kivimaki_2005} remain infectious after beginning antibiotic treatment, and they transmit disease \cite{Siegel_2007}.  Removal (remaining home while infectious) would diminish transmission, though at some inconvenience to the focal infective.  A survey conducted within the last decade suggests that each week nearly $3 \times 10^6$ employees in the U.S. go to work sick \cite{Susser_2016}, fearing lost wages or loss of employment \cite{derigne_2016}.  Tension between pursuit of income and measures intended to curb the spread of infectious disease has become common during pandemic \cite{Maxouris_2020}.

The second observation concerns self-medication in chimpanzees (\emph{Pan troglodytes}).  Chimpanzees consume a diverse plant diet, and at times select plants with antiparasitic properties \cite{Ahoua_2015}.  When infested by intestinal nematodes, a chimpanzee will withdraw from its social group, and while isolated will eat plants with chemical and/or physical characteristics that usually reduce its parasite load \cite{Huffman_1997,Pebsworth_2006}.  As symptoms moderate, the still-parasitized individual can return to the group \cite{Huffman_1996} where its presence may promote transmission of the parasite to other hosts.  Plausibly, self-medication increases survival of the first chimpanzee, and indirectly increases the frequency of parasitism within the group.

\subsection{Summary of predictions}
The results indicate several interrelated predictions, summarized here.
\begin{itemize}
\item The expected count of secondary infections is often a single-peaked function of the time since infection when antibiotic therapy begins.  However, sufficiently strong pathogen self-regulation can imply that $R_0$ increases montonically with time elapsing until therapy begins.
\item Less efficacious antibiotics may increase the expected count of secondary infections beyond the level anticipated without antibiotic intervention.
\item Strong pathogen self-regulation increases the probability that the host remains infectious until therapeutically cured, and decreases the time elapsing between initiation of treatment and cure.
\item Treatment with a less efficacious antibiotic soon after infection can increase the probability of curing the disease, but also can increase the expected count of secondary infections.  However, early treatment with a very strong antibiotic can both increase the likelihood of curing the disease and reduce the count of secondary infections.
\item If hosts are moderately to highly susceptible to infection, duration of the infectious state and the expected count of secondary infections decline as inoculum size increases.
\item When each susceptible individual contacts an infected host at a given stochastic rate, grouping susceptibles increases the variance of the secondary-infection count and, consequently, increases the probability of no new infection.
\end{itemize}

Note that the predictions do not depend on whether removal equates with isolation (usually faster) or host mortality (usually slower).  The next several subsections suggest further questions about the way antibiotics might impact linkage between within-host pathogen growth and among-host transmission.

\subsection{Bacteria}
Genetic resistance to antibiotics, often transmitted \emph{via} plasmids \cite{Lopatkin_2017}, challenges control of bacterial disease \cite{Levin_2014}.  Phenotypic tolerance presents related, intriguing questions \cite{Wiuff_2005}.  Some genetically homogeneous bacterial populations consist of two phenotypes; one grows faster and exhibits antibiotic sensitivity, while the other grows more slowly and can persist after exposure to an antibiotic \cite{Balaban_2004}.  Phenotypes are not fixed; individual lineages may transition between the two forms \cite{Ankomah_2014}.  An antibiotic's effect on densities of the two forms might easily extend the duration of infectiousness, but the probability of transmission, given contact, might decline as the frequency of the persistent type increases.

\subsection{Antibiotic administration}
If an antibiotic is delivered periodically as a pulse, rather than dripped, the therapeutically induced mortality of the pathogen can depend on time since the previous administration \cite{Wiuff_2005}.  Complexity of the impact on the within-host dynamics could then depend on the difference between the antibiotic's decay rate and the pathogen's rate of decline.  Some authors refer to an ``inoculum effect,'' suggesting that antibiotic efficacy can vary inversely with bacterial density.  That is, the \emph{per unit density} bacterial mortality effected by a given antibiotic concentration declines as bacterial density increases \cite{Levin_2010}.

\subsection{Infected host}
This paper neglects immune responses so that the duration of treatment, given cure by the antibiotic, depends explicitly on the antibiotic's efficacy and the age of infection when treatment begins.  Extending the model to incorporate both a constitutive and inducible immune response would be straightforward. Following Hamilton et al. [2008], the constitutive response imposes a constant, density-independent mortality rate on the pathogen.  This response (common to vertebrates and invertebrates) is innately fixed; its effect can be inferred by varying this paper's pathogen growth rate $r$.  Induced immune responses impose density-dependent regulation of pathogen growth; typically, pathogen and induced densities are coupled as a resource-consumer interaction \cite{Pilyugin_2000,Hamilton_2008}.

\subsection{Transmission}
This paper assumes a constant (though probabilistic) rate of infectious contact with susceptible hosts.  The number of contacts available may be limited, so that each transmission event depletes the local-susceptible pool \cite{Dieckmann_2000}.  Regular networks capture this effect for spatially detailed transmission \cite{Caraco_2006}, and networks with a random number of links per host do the same when social preferences drive transmission \cite{vanbaalen_2002}.  For these cases, contact structure of the susceptible population can affect both $R_0$ and the likelihood of pathogen extinction when rare \cite{Duryea_1999,Caillaud_2013}.

Contact avoidance may sometimes be more important than contact depletion \cite{Reluga_2010,Brauer_2011}.  If susceptible hosts recognize correlates of infectiousness, they can avoid individuals or locations where transmission is likely \cite{Meyers_2019}.  Antibiotics might extend the period of infectiousness and, simultaneously, reduce symptom severity.  As a consequence, correlates of infectiousness might be more difficult to detect.

\end{flushleft}

\end{spacing}


\begin{thebibliography}{99}

\bibitem[Ahoua et al. 2015]{Ahoua_2015}
Ahoua, A.R.C., Konan, A.G., Bonfoh, B., Kon\'{e}, M.W., 2015.
Antimicrobial potential of 27 plants consumed by chimpanzees (\emph{Pan troglodytes verus} Blumenbach) in Ivory Coast.
BMCComplementary and Alternative Medicine 15:383 (12 pp).

\bibitem[Ankomah and Levin 2014]{Ankomah_2014}
Ankomah, P., Levin, B.R., 2014.
Exploring the collaboration: antibiotics and the immune response in the treatment of acute, self-limiting infections.
Proc. Natl. Acad. Sci. USA 111:8331--8338. https://doi.org/10.1073/pnas.1400352111.
https://doi.10.1186/s12906.015.0918.7.

\bibitem[Antia et al. 2003]{Antia_2003}
Antia, R., Regoes, R.R., Koella, J.C., Bergstrom, C.T., 2003.
Thde role of evolution in the emergence of infectious diseases.
Nature 426:658--661. https://doi:10.1038/nature02177.

\bibitem[Austin et al. 1998]{Austin_1998}
Austin, D.J., White, N.J., Anderson, R.M., 1998.
The dynamics of drug action on the within-host population growth of infectious agents: melding pharmacokinetics with pathogen population dynamics.
J. Theor. Biol. 194:313--339.

\bibitem[Bailey 1964]{Bailey_1964}
Bailey, N.T.J., 1964.
\emph{The elements of stochastic processes}.
John Wiley \& Sons. New York, USA.

\bibitem[Balaban et al. 2004]{Balaban_2004}
Balaban, N.Q., Marrin, J., Chalt, R., Kowalik, L., Leibler, S., 2004.
Bacterial persistence as a phenotypic switch.
Science 305:1622--1625.  https://doi.org/10.1126/science.1099390.

\bibitem[Brauer 2011]{Brauer_2011}
Brauer, F., 2011.
A simple model for behaviour change in epidemics.
BMC Public Health 11:S3 (5 pp).  https://doi.org/10.1186/1471-2458-11-S1-S3.

\bibitem[Brown et al. 2001]{Brown_2001}
Brown, C.R., Komar, N., Quick, S.B., Sethi, R.A., Panella, N.A., Brown, M.B., Pfeffer, M., 2001.
Arbovirus infection increases with group size.
Proc. R. Soc. London B 268:1833--1849. https://doi:10.1098/rspb.2001.1749.

\bibitem[Bury 1975]{Bury_1975}
Bury, K.V., 1975.
Statistical Models in Applied Science.
John Wiley \& Sons, New York.

\bibitem[Caillaud et al. 2013]{Caillaud_2013}
Caillaud, D., Craft, M.E., Meyers, L.A., 2013.
Epidemiological effects of group size variation in social species.
J. Roy. Soc. Interface 10:20130206. https://doi.org/10.1098/rsif.2013.0206.

\bibitem[Caraco et al. 2016]{Caraco_2016}
Caraco, T., Cizauskas, C.A., Wang, I.-N., 2016.
Environmentally transmitted parasites: Host-jumping in a heterogeneous environment.
J. Theor. Biol. 397:33--42. https://doi.org/10.1016/j.jtbi.2016.02.025

\bibitem[Caraco et al. 2006]{Caraco_2006}
Caraco, T., Glavanakov, S., Li, S., Maniatty, W., Szymanski, B.K., 2006.
Spatially structured superinfection and the evolution of disease virulence.
Theor. Popul. Biol. 69:367--384.  https://doi.org/10.1016/j.tpb.2005.12.004

\bibitem[Caraco et al. 2014]{Caraco_2014}
Caraco, T., Yousefi, A., Wang, I-N. 2014.
Host-jumping, demogrpahic stochasticity and extinction: lytic viruses.
Evol. Ecol. Res. 16:551--568.

\bibitem[Childs et al. 2019]{Childs_2019}
Childs, L.M., El Moustaid, F., Gajewski, Z., Kadelka, S., Nikin-Beers, R., Smith Jr., J.W., Walker, M., Johnson, L.R., 2019.
Linked within-host and between-host models and data for infectious diseases: a systematic review.
PeerJ 7:e7057 (18 pp). https://doi.org/10.7717/peerj.7057.

\bibitem[Chu et al. 2004]{Chu_2004}
Chu, C.-M., Poon, L.L.M., Cheng, V.C.C., Chan, K.-S., Hung, I.F.N., Wong, M.M.L., Chan, K.-W., Leung, W.-S., Tang, B.S.F., Chan, V.L., Ng, W.-L., Sim, T.-C., Ng, P.-W., Law, K.-I., Tse, D.M.W., Peiris, J.S.M, Yuen, K.-Y. 2004.
Initial viral load and the outcomes of SARS.
Canadian Med. Assoc. J. 171:1349--1352. https://doi.org/10.1503/cmaj.1040398

\bibitem[D`Agata et al. 2008]{Dagata_2008}
D`Agata, E.M.C., Dupont-Rouzeyrol, M., Magal, P., Olivier, D., Ruan, S., 2008.
The impact of different antibiotic regimens on the emergence of antimicrobial-resistant bacteria.
PLoS One: e4036 (9 pp). https://doi.org/10.1371/journal.pone.0004036.

\bibitem[D`Argenio et al. 2001]{Dargenio_2001}
D`Argenio, D.A., L.A. Gallagher, C.A. Berg, and C. Manoil. 2001.
\emph{Drosophila} as a model host for \emph{Pseudomonas aeruginosa} infection.
J. Bacteriology 183:1466--1471.

\bibitem[Day et al. 2011]{Day_2011}
Day, T., Alizon, S., Mideo, N., 2011.
Bridging scales in the evolution of infectious disease life histories: theory.
Evol. 65:3448--3461.  https://doi.org/10.1111/j.1558-5646.2011.01394.x.

\bibitem[DeRigne et al. 2016]{derigne_2016}
DeRigne, L., Stoddard, P., Quinn, L., 2016.
Workers without paid sick leave less likely to take time off for illness or injury compared to those with sick leave.
Health Affairs 35:520--527.  https://doi.org/10.1377/hlthaff.2015.0965.

\bibitem[Dieckmann et al. 2000]{Dieckmann_2000}
Dieckmann, U., Law, R., and Metz, J.A.J. (Eds.). 2000.
The Geometry of Ecological Interactions.
Cambridge University Press, Cambridge, UK.

\bibitem[Duryea et al. 1999]{Duryea_1999}
Duryea, M., Caraco, T., Gardner, G., Maniatty, W., Szymanski, B.K., 1999.  Population dispersion and equilibrium infection frequency in a spatial epidemic.
Physica D 132:511 -- 519.  https://doi.org/10.1016/S0167-2789(99)00059-7.

\bibitem[Ebert and Weisser 1997]{Ebert_1997}
Ebert, D., and W.W. Weiser. 1997.
Optimal killing for obligate killers: The evolution of life histories and virulence of semelparous parasites.
Proc. Royal Society B 264:985--991.

\bibitem[Falk et al. 2015]{Falk_2015}
Falk. L., Enger, M., Jense, J.S., 2015.
Time to eradication of \emph{Mycoplasma genitalium} after antibiotic treatment in men and women.
J. Antimicro. Chemotherapy 70:3134-3140. https://doi.org/10.1093/jac/dkv246.

\bibitem[Gama et al. 2012]{Gama_2012}
Gama JA, Abby SS, Vieira-Silva S, Dioisio F, Rocha EP., 2012.
Immune subversion and quorum-sensing shape the variation in infectious dose among bacterial pathogens.
PLoS Pathogens 8:e1002503 (9 pp).

\bibitem[Ganusov and Antia 2003]{Ganusov_2003}
Ganusov, V.V., Antia, R., 2003. Trade-offs and the evolution of virulence of microparasites: do details matter?
Theoret. Popul. Biol. 64, 211–220.  https://doi.org/10.1016/S0040-5809(03)00063-7.

\bibitem[Geli et al. 2012]{Geli_2012}
Geli, P., Laxminarayan, R., Dunne, M., Smith, D.l., 2012.
``One-size-fits-all''? optimizing treatment duration for bacterial infections.
PLoS One 7:e29838 (10 pp). https://doi.org/10.1371/journal.pone.0029838.

\bibitem[Gilchrist and Coombs 2006]{Gilchrist_2006}
Gilchrist, M.A. and Coombs, D. (2006) Evolution of virulence:
interdependence, constraints and selection using nested models.
Theor. Popul. Biol. 69, 145–153

\bibitem[Gilchrist and Sasaki 2002]{Gilchrist_2002}
Gilchrist, M.A. and Sasaki, A. (2002) Modeling host-parasite
coevolution: a nested approach based on mechanistic models. J.
Theor. Biol. 218, 289–308

\bibitem[Gualerzi et al. 2013]{Gualerzi_2013}
Gualerzi, C.O., Brandt, L., Fabbretti, A., Pon, C.L. (eds.) 2013.
Antibiotics: Targets, Mechanisms and Resistance. 549 pp. Wiley-VCH Verlag. Weinhein, Germany. https://doi.org/10.1002/9783527659685.

\bibitem[Hamilton et al. 2008]{Hamilton_2008}
Hamilton, R., Siva-Jothy, M., Boots, M., 2008.
Two arms are better than one: parasite variation leads to combined inducible and constitutive innate immune responses.
Proc. R. Soc. B 275:937--945.  https://doi.org/10.1098/rspb.2007.1574

\bibitem[Haugan et al. 2019]{Haugen_2019}
Haugen, M.S., Hertz, F.B., Charbon, G., Sahin, B., Lobner-Olesen, Frimodt-M{\o}ller, N., 2019.
Growth rate of \emph{Escherichia coli} during human urinary tract infection: implications for antibiotic effect.
Antibiotics 8, 92 (12 pp).  https://doi:10.3390/antibiotics8030092.

\bibitem[Heo et al. 2009]{Heo_2009}
Heo, Y.-J., Y.-R. Lee, H.-H. Jung, J.E. Lee, G.P. Ko, and Y.-H. Cho. 2009.
Antibacterial efficacy of phages against \emph{Pseudomonas aeruginosa} infections in mice and \emph{Drosophila melanogaster}.
Antimicrobial Agents and Chemotherapy 53:2469--2474.

\bibitem[Herrera-Diestra and Meyers 2019]{Meyers_2019}
Herrera-Diestra, J.L., Meyers, L.A., 2019.
Local risk pereception enhances epidemic control.
PLoS ONE 14:e0225576 (15 pp). https://doi.org/10.1371/journal.pone.0225576.

\bibitem[Holdenrieder et al. 2004]{Holdenrieder_2004}
Holdenrieder, O., Pautasso, M., Weisberg, P.J., Lonsdale, D., 2004.
Tree diseases and landscape processes: the challenge of landscape pathology.
Trends Ecol. Evol. 19:446--452. https://doi.org/10.1016/j.tree.2004.06.003.

\bibitem[Huffman et al. 1997]{Huffman_1997}
Huffman, M.A., Gotoh, S., Turner, L.A., Hamai, M., Yoshida, K. 1997.
Seasonal trends in intestinal nematode infection and medicinal plant use among chimpanzees in the Mahale Mountains, Tanzania.
Primates 38:111--125.

\bibitem[Huffman et al. 1996]{Huffman_1996}
Huffman, M.A., Page, J.E., Sukhdeo, M.V.K., Gotoh, S., Kalunde, M.S., Chandrasiri, T., Toswers, G.H.N., 1996.
Leaf-swallowing by chimpanzees: a behavioral adaptation for the control of strongyle nematode infections.
Inter. J. Primatology 17;475--503.

\bibitem[Kaitala et al. 2017]{Kaitala_2017}
Kaitala, V., Roukolainen, L., Holt, R.D., Blackburn, J.K., Merikanto, I., Anttila, J., Laakso, J., 2017.
Population dynamics, invasion, and biological control of environmentally growing opportunistic pathogens, in Hurst, C.J. (Ed.), Modeling the Transmission and Prevention of Infectious Disease. Advances in Environmental Microbiology 4, Springer Intl. Publishing AG, pp. 213--244.

\bibitem[Keebaugh and Schlenke 2014]{Keebaugh_2014}
Keebaugh, E.S. and T.A. Schlenke. 2014.
Insights from natural host-parasite interactions: the \emph{Drosophila} model.
Developmental Comparative Immunology, 42:111--123.

\bibitem[Keeling 1999]{Keeling_1999}
Keeling, M.J., 1999. The effects of local spatial structure on epidemiological invasions.
Proc. R. Soc. London B 266, 859–867.

\bibitem[Keeling and Grenfell 1998]{Keeling_1998}
Keeling, M.J., Grenfell, B.T., 1998.
Effect of variability in infection period on the persistence and spatial spread of infectious diseases.
Math. Biosci. 147:207--226.  https://doi.org/10.1016/S0025-5564(97)00101-6.

\bibitem[Keeling and Rohani 2008]{Keeling_2008}
Keeling, M.J., Rohani, P., 2008.
Modeling Infectious Diseases in Humans and Animals.
Princeton University Press. Prineton, NJ.

\bibitem[Kivimaki et al. 2005]{Kivimaki_2005}
Kivimaki, M., Head, J., Ferrie, J.E., Hemingway, H., Shipley, M.J., Vahtera, J., Marmot, M.G., 2005.
Working while ill as a risk factor for serious coronary events: the Whitehall II study.
Am. J. Public Health 95:98--102. https://doi.org/10.2105/AJPH.2003.035873.

\bibitem[Lahodny et al. 2015]{Lahodny_2015}
Lahodny, G.E., R. Gautam, and R. Ivanek. 2015.
Estimating the probability of an extinction event or major outbreak for an environmentally transmitted infectious disease.
J. Biological Dynamics (S1) 9:128--155.

\bibitem[Levin et al. 2014]{Levin_2014}
Levin, B.R., Baquero, F., Johnsen, P.J., 2014.
A model-guided analysis and perspective on the evolution and epidemiology of antibiotic resistance and its future.
Curr. Opinion Microbiol. 19:83--89. https://doi.org/10.1016/j.mib.2014.06.004.

\bibitem[Levin and Udekwu 2010]{Levin_2010}
Levin, B.R., Udekwu, K.I., 2010.
Population dynamics of antibiotic treatment: a mathematical model and hypotheses for time-kill and continuous-culture experiments.
Antimicro. Agents Chemo. 54:3414--3426. https://doi.org/10.1128/AAC.00381-10.

\bibitem[Lindberg et al. 2018]{Lindberg_2018}
Lindberg, H.M., McKean, K.A., Caraco, T., Wang, I.-N., 2018.
Within-host dynamics and random duration of pathogen infection: implications for between-host transmission.
J. Theor. Biol. 446:137--148. https://doi.org/10.1016/j.jtbi.2018.01.030.

\bibitem[Lopatkin et al. 2017]{Lopatkin_2017}
Lopatkin, A.J., Meredith, H.R., Srimani, J.K., Pfeiffer, C., Durrett, R. You, L., 2017.
Persistence and reversal of plasmid-mediated antibiotic resistance.
Nature Communications 8:1689 (10 pp.).  https://doi.org/10.1038/s41467-017-01532-1.

\bibitem[McManus et al. 2002]{McManus_2002}
McManus, P.S., Stockwell, V.O., Sundin, G.W., Jones, A.L., 2002.
Antibiotic use in plant agriculture.
Annu. Rev. Phytopathol. 40:443–-65 https://doi.org/10.1146/annurev.phyto.40.120301.093927.

\bibitem[Maxouris and Chavez 2020]{Maxouris_2020}
Maxouris, C., Chavez, N., 2020.
Florida will be 'like a house on fire' in weeks with loose coronavirus restrictions, infectious disease expert says.
https://www.cnn.com/2020/10/09/health/us-coronavirus-friday/index.html (accesssed 31 October 2020).

\bibitem[Medzhitov et al. 2012]{Medzhitov_2012}
Medzhitov, R., Schneider, D.S., Soares, M.P., 2012.
Disease tolerance as a defense strategy.
Science 335:936--941.  https://doi.org/10.1126/science.1214935.

\bibitem[Mideo et al. 2008]{Mideo_2008}
Mideo, N., Alizon, S., Day, T., 2008.
Linking within- and between-host dynamics in the evolutionary epidemiology of infectious diseases.
Trends Ecol. Evol. 23:511--517. https://doi.org/10.1016/j.tree.2008.05.009.

\bibitem[Missov and Lenart 2013]{Missov_2013}
Missov, T.I., Lenart, A. 2013.
Gompertz-Makeham life expectancies: expressions and applications.
Theor. Pop. Biol. 90:29--35.  http://dx.doi.org/10.1016/j.tpb.2013.09.013.

\bibitem[Moon 2019]{Moon_2019}
Moon, M.-S., 2019.
Essential basic bacteriology in managing musculoarticuloskeletal infection: Bacterial anatomy, their behavior, host phagocytic activity,
immune system, nutrition, and antibiotics.
Asian Spine J. 13:343--356.  https://doi.org/10.31616/asj.2017.0239.

\bibitem[Mueller et al. 2004]{Mueller_2004}
Mueller, M., de la Pe\~{n}a, A., Derendorf, H., 2004.
Issues in pharmacokinetics and pharmacodynamics of anti-infective agents: kill curves versus MIC.
Antimicro. Agents Chemo. 48:369--377.  https://doi.org/10.1128/AAC.48.2.369–377.2004.

\bibitem[Mulcahy et al. 2011]{Mulcahy_2011}
Mulcahy, H., C.D. Sibley, M.G. Surette, and S. Lewenza. 2011.
\emph{Drosophila melanogaster} as an animal model for the study of \emph{Pseudomonas aeruginosa} biofilm infections \emph{in vivo}.
PLoS Pathogens 7:e1002299 (14 pp).

\bibitem[O`Loughlin et al. 2013]{Oloughlin_2013}
O`Loughlin, C.T., L.C. Miller, A. Siryaporn, K. Drescher, M.F. Semmelhack, and B.L. Bassler. 2013.
A quorum-sensing blocks \emph{Pseudomonas aeruginosa} virulence and biofilm formation.
Proc. Natl. Acad. Science USA 110:17981--17986.

\bibitem[Pebsworth et al. 2006]{Pebsworth_2006}
Pebsworth, P., Krief, S., Huffman, M.A., 2006.
The role of diet in self-medication among chimpanzees in the Sonso and Kanyawara  comunities, Uganda, in Newton-Fisher, N.E., Norman, H., Reynolds, W., Paterson, J.D. (Eds.), Primates of Western Uganda (Pp 105--133). Springer, New York.

\bibitem[Pilyugin and Antia 2000]{Pilyugin_2000}
Pilyugin, S.S., Antia, R., 2000.
Modeling immune responses with handling time.
Bull. Math. Biol. 62:869--890.  https://doi.org/10.1006/bulm.2000.0181.

\bibitem[Read et al. 2011]{Read_2011}
Read, A.F., Day, T., Huijben, S., 2011.
The evolution of drug resistance and the curious orthodoxy of aggessive chemotherapy.
Proc. Natl. Acad. Science USA 108:10871--10877.  www.pnas.org/cgi/doi/10.1073/pnas.1100299108.

\bibitem[Regoes et al. 2004]{Regoes_2004}
Regoes, R.R., Wiuff, C., Zappala, R.M., Garner, K.M., Baquero, F., Levin, B.R., 2004.
Pharmacodynamic functions: a multiparameter approach to the design of antibiotic treatment regimens.
Antimicro. Agents Chemo. 48:3670--3676.  https://doi.org/10.1128/AAC.48.10.3670–3676.2004.

\bibitem[Reluga 2010]{Reluga_2010}
Reluga, T.C., 2010.
Game theory of social distancing in response to an epidemic.
PLoS Comput. Biol. 6:e1000793 (9 pp).  https://doi.org/10.1371/journal.pcbi.1000793.

\bibitem[Ross 1983]{Ross_1983}
Ross, S.M., 1983.
Stochastic Processes.
John Wiley \& Sons, New York.

\bibitem[Schmid-Hempel and Frank 2007]{Schmid_2007}
Schmid-Hempel P, Frank SA., 2007.
Pathogenesis, virulence, and infective dose.
PLoS Pathogens 3:e147 (2 pp).

\bibitem[Siegel et al. 2007]{Siegel_2007}
Siegel, J.D., Rhinehart, E., Jackson, M., Chiarello, L., Healthcare Infection Control Practices Advisory Committee., 2007.
Guideline for isolation precautions: Preventing transmission of infectious agents in healthcare settings,
https://www.cdc.gov/infectioncontrol/guidelines/isolation/index.html (accessed 12 October 2019).

\bibitem[Steinmeyer et al. 2010]{Steinmeyer_2010}
Steinmeyer, S.H., Wilke, C.O., Pepin, K.M., 2010.
Methods of modelling viral disease dynamics across the within- and between-host scales: the impact of viral dose on host population immunity.
Phil. Trans. R. Soc. B 365:1931--1941.  https://doi.org/10.1098/rstb.2010.0065.

\bibitem[Strachan et al. 2005]{Strachan_2005}
Strachan, N.J.C., Doyle, M.P., Kasuga, F., Rotariu, O., Ogden, I.D., 2005.
Dose response modelling of \emph{Escherichia coli} O157 incorporating data from foodborne and environmental outbreaks.
Int. J. Food Microbiol. 103:35--47.  https://doi.org/10.1016/j.ijfoodmicro.2004.11.023.

\bibitem[Strauss et al. 2019]{Strauss_2019}
Strauss, A.T., Shoemaker, L.G., Seabloom, E.W., Borer, E.T., 2019.
Cross-scale dynamics in community and disease ecology: relative timescales shape the community ecology of pathogens.
Ecology:e02836. https://doi.org/10.1002/ecy.2836.

\bibitem[Susser and Ziebarth 2016]{Susser_2016}
Susser, P., Ziebarth, H.R., 2016.
Profiling the U.S. sick leave landscape: presenteeism among females.
Health Services Research 51:2305-2317.  https://doi.org/10.1111/1475-6733.12471.

\bibitem[Tenuis et al. 1996]{Tenuis_1996}
Tenuis, P.F.M., van der Heijden, O.G., van der Giessen, J.W.B., Havelaar, A.H., 1996.
The dose-response relation in human volunteers for gastro-intestinal pathogens.
National Institute of Public Health and the Environment. Bilthoven, The Netherlands.

\bibitem[Tuomanen et al. 1986]{Tuomanen_1986}
Tuomanen, E., Cozens, R., Tosch, W., Zak, O., Tomasz, A., 1986.
The rate of killing of \emph{Escherichia coli} by $\beta-$lactam antibiotics is strictly proportional to the rate of bacterial growth.
J Gen. Microbio. 132:1297--1304.

\bibitem[Turner et al. 2008]{Turner_2008}
Turner, J., Bowers, R.G., Clancy, O., Behnke, M.C., Christley, R.M., 2008.
A network model of \emph{E. coli} O157 transmission within a typical UK dairy herd: the effect of heterogeneity and clustering on the prevalence of infection.
J. Theor. Biol. 254:45--554. https://doi.org/10.1016/jtbi.2008.05.007.

\bibitem[van Baalen 2002]{vanbaalen_2002}
van Baalen, M., 2002.
Contact networks and the evolution of virulence, in Dieckmann, U., Metz, J.A.J., Sabelis, M.W., Sigmund, K., Law,
R., Metz, H. (Eds.), Adaptive Dynamics of Infectious Diseases: In Pursuit of Virulence Management.  Cambridge University Press, Cambridge, pp. 85--103.

\bibitem[VanderWall and Ezenwa 2016]{Vanderwaal_2016}
VanderWaal, K.L., Ezenwa, V.O., 2016.
Heterogeneity in pathogen transmission: mechanisms and methodology.
Funct. Ecol. 30:1607--1622.  https://doi.org/10.1111/1365-2435.12645.

\bibitem[White et al. 2012]{White_2012}
White, S.M., J.P. Burden, P.K. Maini, and R.S. Hails. 2012.
Modelling the within-host growth of viral infections in insects.
J. Theoretical Biology 312:34--43.

\bibitem[Whittle 1955]{Whittle_1955}
Whittle, P. 1955.
The outcome of a stochastic epidemic: a note on Bailey`s paper.
Biometrika 42:116--122.

\bibitem[Wiuff et al. 2005]{Wiuff_2005}
Wiuff, C., Zappala, R.M., Regoes, R.R., Garner, K.N., Baquero, F., Levin, B.R., 2005.
Phenotypic tolerance: antibiotic enrichment of noninherited resistance in bacterial populations.
Antimicro. Agents Chemo. 49:1483--1494. https://doi.org/10.1128/AAC.49.4.1483-1494.2005.

\end{thebibliography}
\end{document}